\documentclass[pdflatex,sn-mathphys-num,Numbered]{sn-jnl}

\usepackage{graphicx}%
\usepackage{multirow}%
\usepackage{amsmath,amssymb,amsfonts}%
\usepackage{amsthm}%
\usepackage{mathrsfs}%
\usepackage[title]{appendix}%
\usepackage{xcolor}%
\usepackage{textcomp}%
\usepackage{manyfoot}%
\usepackage{booktabs}%
\usepackage{algorithm}%
\usepackage{algorithmicx}%
\usepackage{algpseudocode}%
\usepackage{listings}%
\usepackage{siunitx}
\usepackage{lineno}
\usepackage{stmaryrd}

\newcommand{\bq}{\begin{equation}}
\newcommand{\eq}{\end{equation}}
\newcommand{\bqa}{\begin{eqnarray}}
\newcommand{\eqa}{\end{eqnarray}}
\newcommand{\baa}[1]{\begin{array}{#1}}
\newcommand{\eaa}{\end{array}}

\def\c2{\chi^2}

\def\gg{\gamma\gamma}

\def\gggg{\gamma\gamma\rightarrow\gamma\gamma}
\def\gggz{\gamma\gamma\rightarrow\gamma Z}
\def\ggzz{\gamma\gamma\rightarrow ZZ}

\begin{document}
\title{A realistic photon spectra in polarized $\gg$ processes in \texttt{SANCphot}.}


\author[1,3]{\fnm{Sergey} \spfx{G.} \sur{Bondarenko}}
\author[2,3]{\fnm{Aidos} \sur{Issadykov}}
\author[1]{\fnm{Lidia} \spfx{V.} \sur{Kalinovskaya}}
\author*[1]{\fnm{Andrey} \spfx{A.} \sur{Sapronov}}\email{andrey.a.sapronov@gmail.com}

\affil[1]{\orgdiv{Dzhelepov Laboratory for Nuclear Problems}, \orgname{JINR}, \orgaddress{\street{Joliot-Curie 6}, \postcode{RU-141980} \city{Dubna}, \country{Russia}}}
\affil[2]{\orgdiv{Bogoliubov Laboratory of  Theoretical Physics}, \orgname{JINR}, \orgaddress{\street{Joliot-Curie 6}, \postcode{RU-141980} \city{Dubna}, \country{Russia}}}
\affil[3]{\orgdiv{The Institute of Nuclear Physics}, \orgname{Agency of the Republic of Kazakhstan for Atomic Energy},  \orgaddress{\street{Ibragimov Street 1},\postcode{KZ-050032}, \city{Almaty},  \country{Kazakhstan}}}


\abstract{This work presents an approach to improve the precision of polarized photon-photon 
collisions simulation implemented in the \texttt{SANCphot} package. The basic
linear Compton approximation of the incoming photon spectrum is extended to 
a general energy distribution and a realistic description of circular or 
linear polarizations as expected to be seen at photon-photon colliders.}


\maketitle

\section{Introduction} \label{sec:intro}
Photon colliders, envisioned as potential future high-energy physics research
instruments, offer a unique opportunity to explore electroweak and
beyond-the-Standard-Model (BSM) physics in unprecedented precision. While the
phenomena of photon-photon collisions were previously observed as a side effect
in electron-positron and heavy ion collisions, a dedicated machine would open up
new avenues for precision electroweak physics. An overview presented
in~\cite{ECFADESYPhotonColliderWorkingGroup:2001ikq} covers a wide range of
topics dedicated to the development of a photon-photon collider, including
physics program, photon beam generation, and effects in the interaction region.
Theoretical calculations and simulation codes accompany these developments,
gradually improving our understanding and correcting expectations from these
intriguing projects.

In this paper, we investigate the realistic simulation of polarized
photon-photon collisions at photon colliders by coupling the detailed photon
beam simulation and the theoretical calculations for the $\gg$ processes. A
precise description of the $\gg$ processes that considers initial photon
polarization is useful in a couple of aspects. When studied in angular
distributions, polarization effects of the Z-boson in processes $\gggz$ and
$\ggzz$ provide sensitivity to potential new physics beyond the Standard Model.
Anomalous quartic gauge couplings and extra dimensions are another kind of
phenomena, where accurate simulation of polarized photon-photon collisions can
improve background estimation and lead to better
constraints~\cite{Brodsky:1994nf} in these domains.

Previously, several remarkable works have been devoted to the study of $\gg$
processes, in particular, light-by-light scattering (LbL), taking into account
radiative corrections~\cite{Karplus:1950zz,Jikia:1993tc,Bohm:1994sf,
Gounaris:1998qk,Bern:2001dg}. The process of $\gamma\gamma \to \gamma Z$ was
reviewed in~\cite{Gounaris:1999ux}. A calculation of ZZ production cross-section
components ($\sigma_0, \sigma_{22},$ etc.) was presented and available as a
public code~\cite{Diakonidis:2006cv}. Processes $\gggg$ and $\gggz$ with
polarized incoming beams at CLIC are investigated in terms of anomalous
couplings~\cite{Inan:2020gfu, Inan:2021pbx}.

The paper is composed as follows. A short preliminary word on the current status
of the photon-photon collider is given in the Introduction~\ref{sec:intro}. We review
how photon-photon collisions are treated in the SANCphot code in
Section~\ref{sec:sanc_gg}. Then, in Section~\ref{sec:pc_spectrum}, we overview
challenges occurring when trying to give a precise description of photon spectra
generated at the photon-photon colliders by the Compton scattering. Our method
of supplying a realistic photon spectrum to the SANCphot calculation is described
in Section~\ref{sec:sanc_cain}, and numeric results of this work are presented
in Section~\ref{sec:num_results}. The summary in Section~\ref{sec:summary} gives a brief
overview of our studies.

\section{Photon-photon collisions in SANC}
\label{sec:sanc_gg}
The cross section of photon-photon collisions implemented in the SANCphot
tool~\cite{Bondarenko:2022ddm} follows a recipe described
in~\cite{Gounaris:1999gh} using FORTRAN modules introduced
in~\cite{Bardin:2009gq, Bardin:2012mc, Bardin:2017qqw}. The total $\gg$ process
cross section is represented as a convolution of its components in different
helicity combinations $d\bar{\sigma}_{ij}/d\cos\vartheta^*$ and averaged Stokes
parameters defining the density matrix of the back-scattered photon 
$\langle\xi_i \xi_j^\prime \rangle$:
\begin{align}
\begin{split}
{\frac{d\sigma}{dx_1dx_2 d\cos\vartheta^*}}=&
{\frac{d\bar L_{\gamma\gamma}}{ dx_1dx_2}} \Bigg \{
{\frac{d\bar{\sigma}_0}{d\cos\vartheta^*}}
+\langle \xi_2 \xi_2^\prime \rangle
{\frac{d\bar{\sigma}_{22}}{d\cos\vartheta^*}} \\
& +[\langle\xi_3\rangle\cos2\phi+\langle\xi_3^ \prime\rangle\cos2\phi^\prime]
{\frac{d\bar{\sigma}_{3}}{d\cos\vartheta^*}} \\
&+\langle\xi_3 \xi_3^\prime\rangle[
{\frac{d\bar{\sigma}_{33}}{d\cos\vartheta^*}}
\cos2(\phi+\phi^\prime) \\
&+{\frac{d\bar{\sigma}^\prime_{33}}{d\cos\vartheta^*}}
\cos2(\phi- \phi^\prime)] \\
&+[\langle\xi_2 \xi_3^\prime\rangle\sin2 \phi^\prime \\
&-\langle\xi_3 \xi_2^\prime\rangle\sin2\phi]
{\frac{d\bar{\sigma}_{23}}{d\cos\vartheta^*}} \Bigg \} \ , 
\label{sigpol}
\end{split}
\end{align}
In this definition, the $\vartheta^*$ is the scattering angle of $\gg$ collision
products measured in the rest frame of the incoming particles, and
$x_1, x_2$ --- a fractions of the electron's energy
transferred to the back-scattered photon. Further, $\frac{d\bar{L}_{\gg}}{dx_1dx_2}$
denotes the photon-photon luminosity per unit $e^-e^-$ flux, considering the
linear collider operating in $\gg$ mode, and the Stocks parameters $\xi_2,
\xi_3$ are defined via the photon and electron polarization functions as
described in~\cite{Ginzburg:1981vm},~\cite{Ginzburg:1982yr}. The
$\frac{d\bar{L}_{\gg}}{dx_1dx_2}$ factor, consequently, depends on the energy
distribution of the back-scattered photons and can be evaluated analytically
using the linear Compton approximation (LCA). 

\section{Photon spectrum in photon colliders}
\label{sec:pc_spectrum}
The linear Compton approximation has several drawbacks when utilized in the
simulation of the $\gg$ collisions in the photon-photon
collider~\cite{Telnov:1989sd}. Most of the inaccuracy is attributed to the
process of electron beam conversion to a high-energy photonic beam. Here, the
oncoming laser beam has high but nonuniform intensity, and the electron bunches have
distorted geometry, nonuniform electron density, and a strong electromagnetic field.
More difficulties are added by the fact that the particle momenta are not
perfectly aligned with the beam axis and the presence of an external magnetic
field at the interaction point.

Several works have been attributed to improving the spectrum of the photons
generated via electron energy transfer. To our knowledge, the said issues are
best addressed in the photon collider simulation code CAIN~\cite{Chen:1994jt}. The
substantial improvements are achieved by introducing full treatment of nonlinear
and polarized Compton scattering, a detailed description of the space-time
structure of the electron and laser beams, complete kinematics, including angular
distributions, and interplay with beamstrahlung and other background processes.
Other efforts to improve photon spectrum approximation are presented in
works~\cite{Galynskii:2000fk, Ivanov:2004fi} which contribute to a detailed
description of nonlinear effects and a complete description of the photon
polarization in Compton back-scattering.

Public codes suitable for photon spectrum simulation include
CompAZ~\cite{Zarnecki:2002qr} and CIRCE1~\cite{Ohl:1996fi}, both based on the
creation of explicitly parameterized distributions. These tools are focused on
approximating the high-energy regions of the photon spectra.
CIRCE2~\cite{Ohl:2002ff} features correlated double-beam spectra based on
tabulated energy distributions built from the event records generated in
advance. 

\section{Coupling CAIN and SANCphot}
\label{sec:sanc_cain}
In order to supply the realistic photon spectra obtained with the CAIN
simulation, including the polarization information, to the Monte Carlo
integration routines implemented in SANCphot, we use an approach similar to
CIRCE2~\cite{Ohl:2002ff}. The event records, generated with a realistic laser
and oncoming electron beam parameters, along with a non-linear Compton process
expressions, are aggregated into piecewise continuous functions, representing
approximate distributions of spectra and linear and circular polarizations. 
\begin{align}
f_L(x) = \sum_{i=0}^{n}\bar{L}_i\chi_i(x), \; \chi_i(x) = 
\begin{cases}
     1\;\mathrm{if}\; \left\llbracket x/n \right\rrbracket = 1\\
     0\;\mathrm{if}\; \left\llbracket x/n \right\rrbracket \ne 1\\
\end{cases}
\end{align}
where $f_L$ is a piecewise approximation of the luminosity, $\chi_i(x)$ is an
indicator function, $L_i$ is the aggregated $L_{\gamma\gamma}$ value in bin $i$.
Similar definitions are used for approximations of the Stokes parameters:
$f_{\xi_2}(x)$ and $f_{\xi_3}(x)$.  The cubic interpolation of these functions
serves as an input spectrum to \texttt{SANCphot}'s integration procedures in the
same manner as analytic expressions of linear Compton.

As an example of a realistic photon spectrum simulation, the parameters of the
CLIC accelerator were chosen to configure the CAIN code: number of electrons per
bunch $N = 4\times10^9$, the RMS (root-mean-square) bunch length in the longitudinal direction
$\sigma_z = 0.05$~mm, normalized emittance in the y direction
$\gamma\varepsilon_{y} = 1\times10^{-7}\mathrm{m}\cdot \mathrm{rad} $ and in x
direction $1.9\times10^{-6}$ and $1.5\times10^{-6}\mathrm{m}\cdot \mathrm{rad}$
for 500 and 1000~GeV respectively, beta functions $\beta_{x,y} = 2/0.1$~mm and
the transverse beam sizes $\sigma_{x,y} = 88/4.5$~nm for 500~GeV and $55/3.2$~nm
for 1000~GeV. The following laser parameters were used: pulse energy $U_L = 0.8$~J,
pulse duration $\sigma_t = 1$~ps, Rayleigh lengths $z_{Rx} = z_{Ry} = 0.1$~mm, crossing
angle $\phi_c = 14$~mrad. These settings correspond to beam waist $\sigma_0=\qty{2.79}{\micro\metre}$.

The polarizations of the back-scattered photons in CAIN were controlled via
Stokes parameters $(\xi_1, \xi_2, \xi_3)$. Simulation of non-linear Compton
scattering is controlled via the number of absorbed laser photons $n_{ph}$
parameter in CAIN. However, the support for electron transverse polarization is limited only
to linear Compton approximation, and for values $n_{ph} > 0$ the laser photon
polarization is narrowed to the circular one (i.e. $\xi_1 = \xi_3 =
0,\,\xi_2=\pm 1 $). The difference between $n_{ph} = 0$ and $n_{ph} = 1$
setting is that the former uses linear Compton formulas, which are inapplicable
in strong laser fields. The strength of a laser
electromagnetic field is characterized by a parameter 
$\xi^2 = 2n_\gamma r_e^2 \lambda/\alpha$, where $n_\gamma$ is the density of laser photons, 
$\lambda$ is the laser wavelength, $\alpha=e^2/\hbar c = 1/137$, $r_e$ - electron radius
\cite{ECFADESYPhotonColliderWorkingGroup:2001ikq, Ivanov:2004fi}.
At $\xi < O(1)$, the non-linear
Compton expression requires expansion in terms of $n_{ph}$ to provide a better
description. For the laser configuration provided in the above paragraph $\xi^2 \approx 0.3$
and the number of photons per electron is $k = 4.1$ for
500~GeV and $5.6$ for 1000~GeV, which justifies modeling the process as multiple
Compton scattering.
%
\begin{figure*}
  \begin{center}
\includegraphics[width=0.3\textwidth]{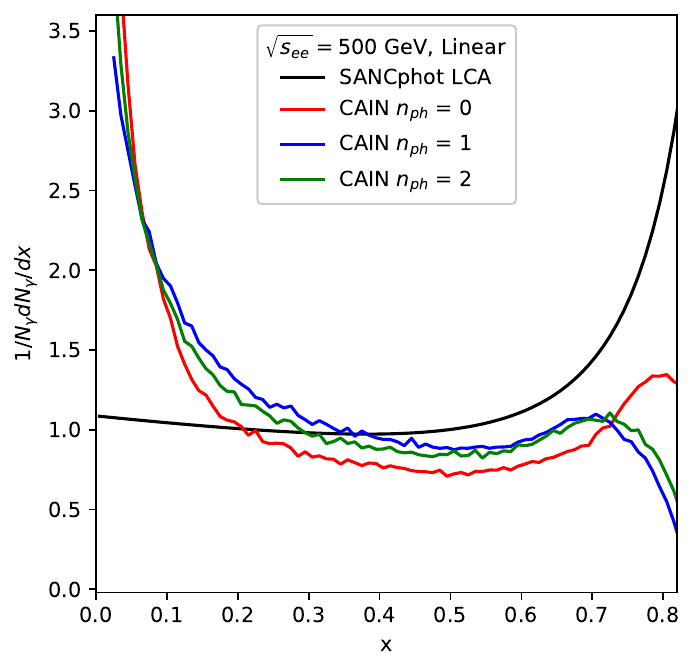}
\includegraphics[width=0.3\textwidth]{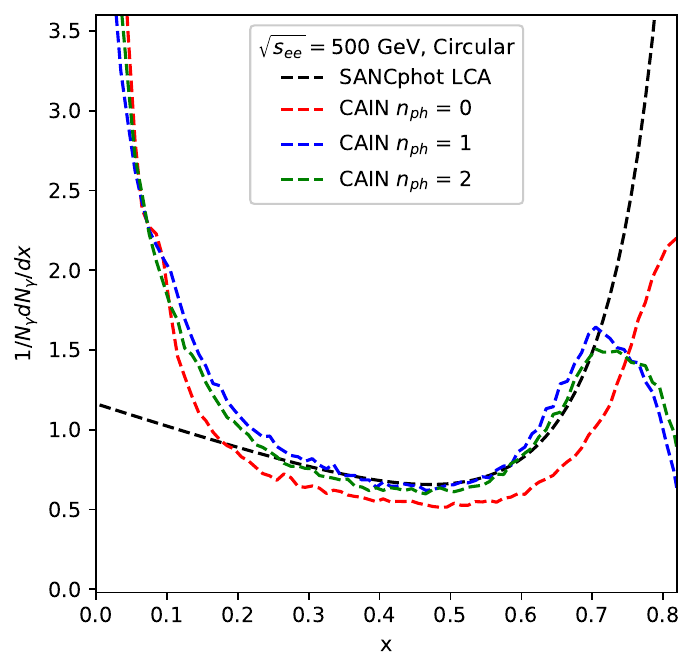} \\
\includegraphics[width=0.303\textwidth]{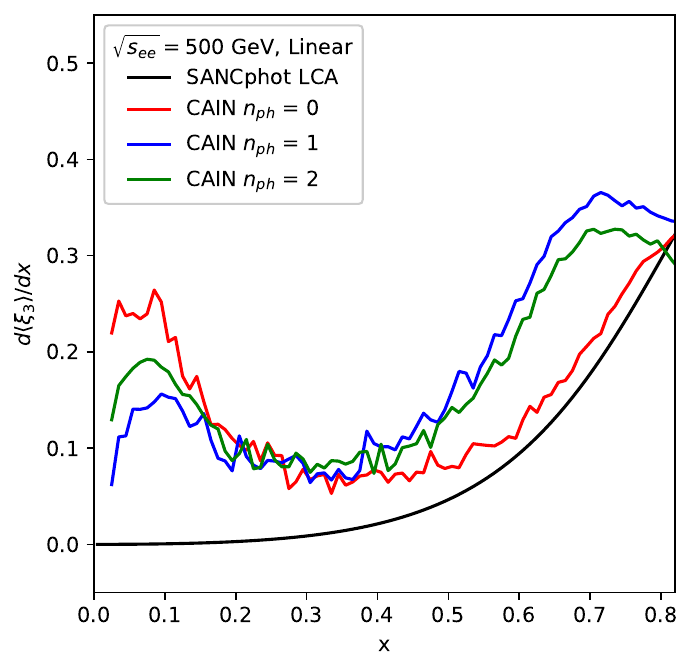}
\includegraphics[width=0.315\textwidth]{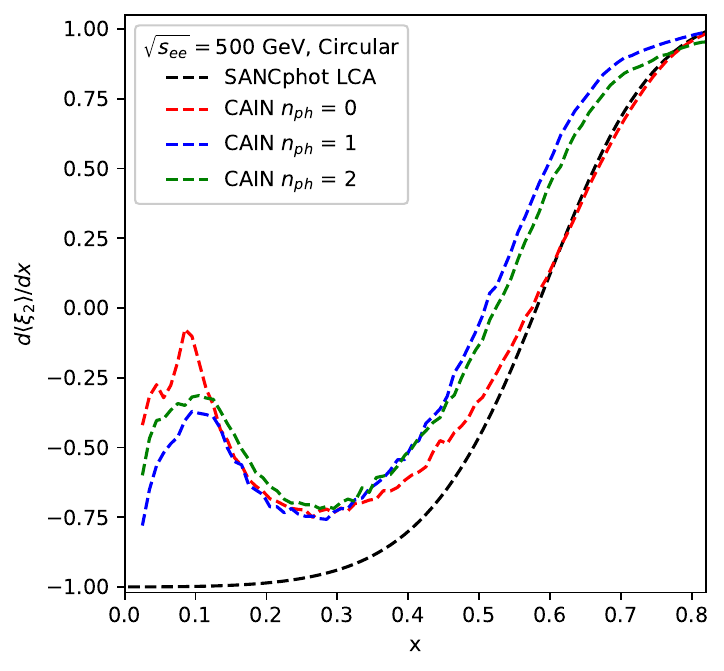} \\
  \end{center}
  \caption{A comparison of normalized photon energy spectra $\frac{1}{N_\gamma} \frac{dN_\gamma}{dx}$ and cross section polarizing 
  coefficients $d\langle\xi2\rangle/dx$ and $d\langle\xi3\rangle/dx$ depending on
  the fraction of transferred energy $x$. The left and right columns correspond 
  to linear and circular laser polarizations respectively.\label{fig:spectra}}
\end{figure*}

In the present study, we use two distinct polarization configurations:
\begin{itemize}
\item linear  : $P_e = P_e' = 0,   P_{\gamma} = P_{\gamma}' = 0,  P_t = P_t' = 1, \phi=\pi/2$
\item circular  : $P_e = P_e' = 0.8, P_{\gamma} = P_{\gamma}' = -1, P_t = P_t' = 0$
\end{itemize}
Here $P_e$ is an electron longitudinal polarization, $P_{\gamma}$ describes the average
helicity of the laser photon, $P_t$ denotes maximum average transverse polarization along
a direction determined by the azimuthal angle $\phi$. This angle is determined
with respect to the back-scattered photon momentum.

Figure~\ref{fig:spectra} shows photon energy spectra and polarization at different levels
of approximation: LCA calculated with \texttt{SANCphot} (black lines), linear and non-linear Compton
approximations calculated by CAIN with $n_{ph} = $ 0 (red), 1 (blue), and 2 (green) with
realistic beam geometry. The red curve shows effects related to the multiple photon scattered
per electron, the blue curves depict engagement of non-linear calculations, giving
significant shift towards lower energy regions, but the green curves corresponding to higher
harmonics show slight shift towards higher energy regions which complies with Fig.2 of~\cite{Ivanov:2004fi}.

\section{Numerical results}
\label{sec:num_results}

The numerical validation was performed in the $\alpha(0)$ EW scheme using the setup given below. For ease
of validation, we keep the electroweak parameter values the same as in~\cite{Bondarenko:2022ddm}:\\
\begin{tabular}{ll}
$\alpha =1/137.035990996$,      & $G_{F} = 1.13024\times 10^{-5}\,\mathrm{GeV}^{-2}$,  \\
$M_{W}= 80.45149$\,GeV,           & $M_{Z}= 91.18670$\,GeV,            \\  
$M_{H}= 125$\,GeV,              &                                   \\
$m_e = 0.51099907$\,MeV,   & $m_\mu = 0.105658389$\,GeV,       \\
$m_\tau = 1.77705$\,GeV,        &                   \\
$m_u = 0.062$\,GeV,     & $m_d = 0.083$\,GeV,           \\
$m_c = 1.5$\,GeV,       & $m_s = 0.215$\,GeV,           \\
$m_t = 173.8$\,GeV,             & $m_b = 4.7 $\,GeV.           \\
\end{tabular} \\
Initial photon spectra were approximated
using several methods: (i) analytic linear Compton approximation implemented in
\texttt{SANCphot}, (ii) photon energy distribution representing linear Compton
spectrum, (iii) realistic spectra obtained from CAIN photon collider simulation
package. The initial electron energy was picked from 500~GeV and 1~TeV, and the
laser beam energy irradiating the electrons was set to 1.26120984~eV, assuming
that it provides an optimal conversion ratio for the back-scattered photons with
parameter $x_0 = 4.83$ (at electron beam energy 500~GeV). The analytic
expressions for the 4-boson interaction diverge at large scattering angles, making
the numeric calculation unstable. Therefore, the phase space is constrained by
$30^{\circ} < \theta < 150^{\circ}$. Also the final state invariant mass for the
processes $\gggg,\, \gggz,\, \ggzz$ was constrained to be above 20, 100 and 200~GeV
correspondingly to ignore kinematically forbidden regions.

\begin{table}[!h]
\centering
\begin{tabular}{llll}
\toprule
   LCA&  SANCphot & SANCphot & CAIN   \\
 spectrum &  analytic & piecewise & piecewise   \\
\midrule
linear  & 8.2019(4)&8.2023(4)&8.4657(4) \\
circular  & 8.8809(4)&8.8811(4)&9.3032(5) \\
\bottomrule

\end{tabular}
  \caption{Integrated cross sections $\sigma(\gggg)$~[fb] for linear Compton approximation spectra.\label{tab:lca_xs_gggg}}
\end{table}

\begin{figure*}
  \begin{center}
\includegraphics[width=0.31\textwidth]{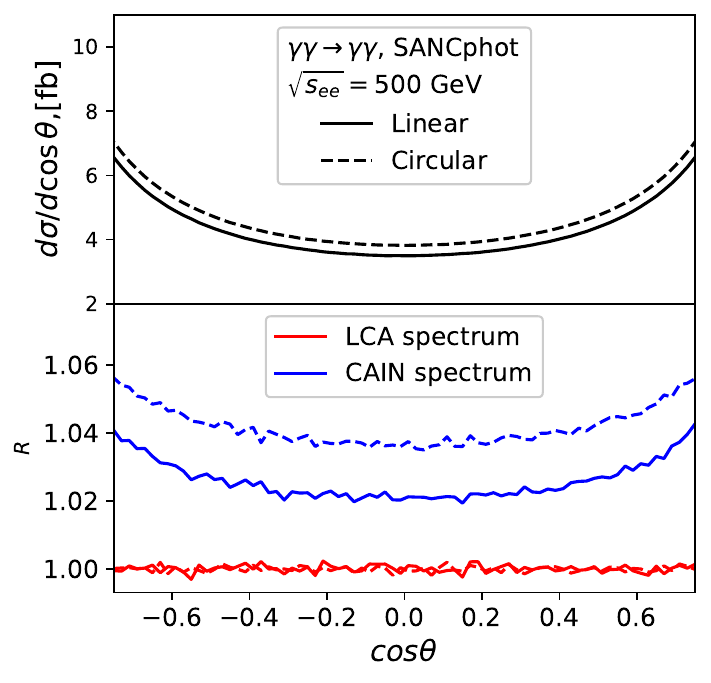}
\includegraphics[width=0.315\textwidth]{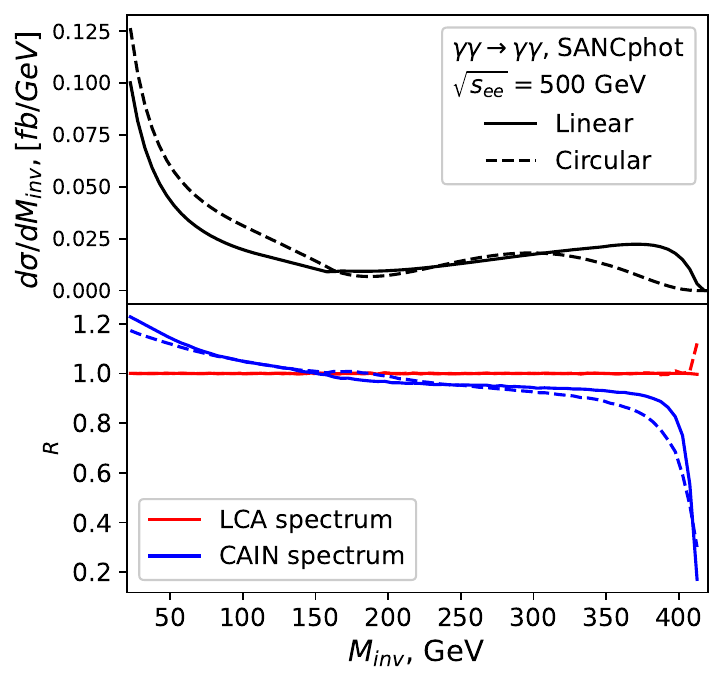}
\includegraphics[width=0.315\textwidth]{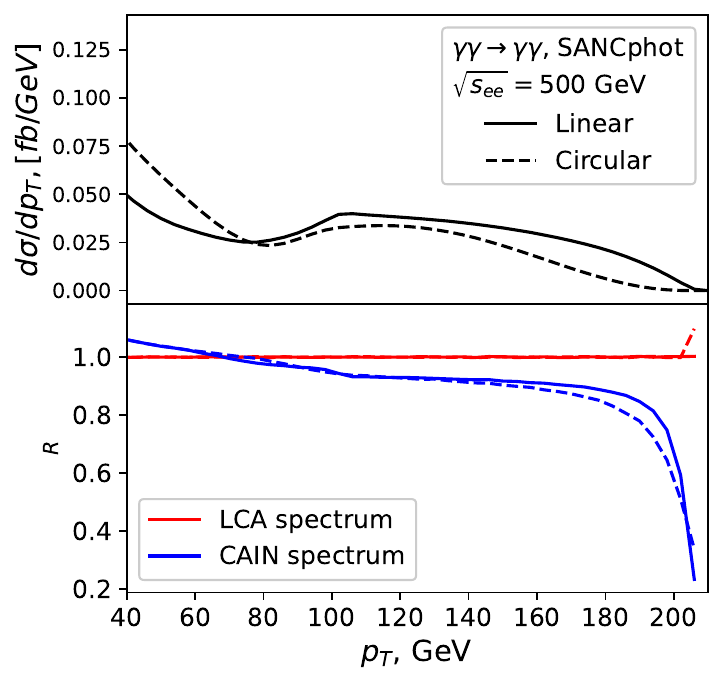} \\
  \end{center}
  \caption{The kinematic distributions for $\gggg$ process at 500GeV with linear Compton approximation. 
  Upper plots are based on the analytic expressions from \texttt{SANCphot}. Lower plots give the ratios of
  distributions built using piecewise approximation of \texttt{SANCphot} (red) and CAIN (blue) spectra
  to the analytic values.\label{fig:kin_dist_spectra}}
\end{figure*}

We perform several checks to validate the simulation of photon-photon collisions
with arbitrary initial photon spectra and polarization distributions. Firstly,
we compare the total cross sections and kinematic
distributions in azimuthal angle, invariant mass, and transverse momentum
computed using analytic expressions for linear Compton approximation and
those calculated via pre-generated LCA photon spectra for transverse and circular
photon polarizations shown in Figure~\ref{fig:kin_dist_spectra}. In these cases, the photon
conversion point coincides with the central interaction point. Second, we
compare with the cross sections based on the spectra obtained with the
photon-photon collision simulation by CAIN to validate against established
simulation package and justify its further usage for realistic spectra
simulation. CAIN simulates the realistic beam shapes and angular distributions,
therefore, despite its settings being set to linear Compton $(n_{ph} =0)$, the
spatial effects produce visible differences from the pure Compton scattering.

In addition, Table~\ref{tab:lca_xs_gggg} shows numeric comparison of three
different sources of photon spectra, when using the analytic version from SANCphot, 
piecewise spectrum description of the analytic expression, and LCA obtained with CAIN 
minimizing beam geometry effects. For the latter the laser parameters were set to $z_R = 1$~mm,
$\sigma_t = 100$~ps, which corresponds to $\xi^2 \approx 3\times10^{-4}$, ensuring the
linear mode for Compton back-scattering. We see that the piecewise description of 
pure LCA spectrum is very close to the analytic one.

For a more precise photon beam simulation in the beam parameters similar to the
CLIC accelerator, we consider the invariant mass and transverse momentum
distributions plotted on the Figures~\ref{fig:kin_m34} and~\ref{fig:kin_pt3}. 
The plots show differential cross sections calculated using
analytic expressions for linear Compton approximation (black lines), linear
Compton with realistic beam geometry simulated in CAIN (red lines), and taking
into account non-linear Compton effects together with a realistic beam geometry
simulated in CAIN for $n_{ph} = 1$ (blue) and $n_{ph} = 2$ (green). The solid
and the dashed lines correspond to circular and linear photon polarization,
respectively.
\begin{figure*}
  \begin{center}
\includegraphics[width=0.315\textwidth]{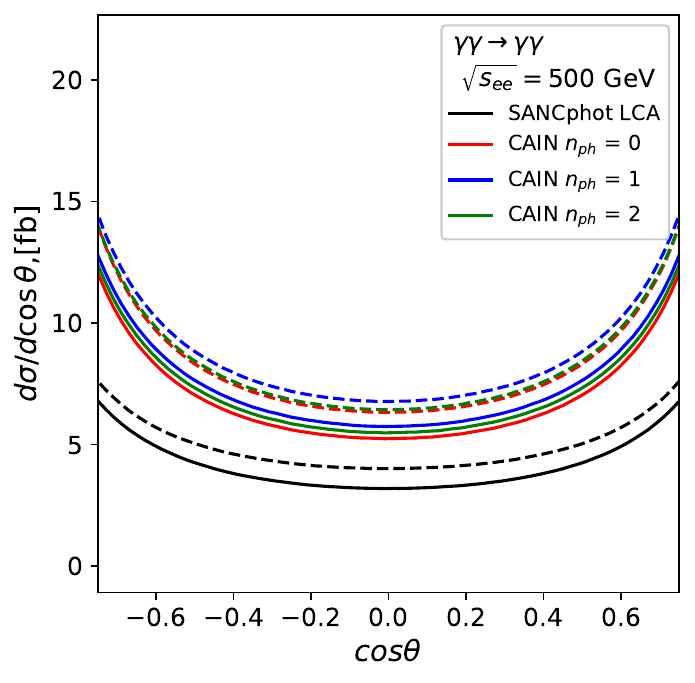}
\includegraphics[width=0.315\textwidth]{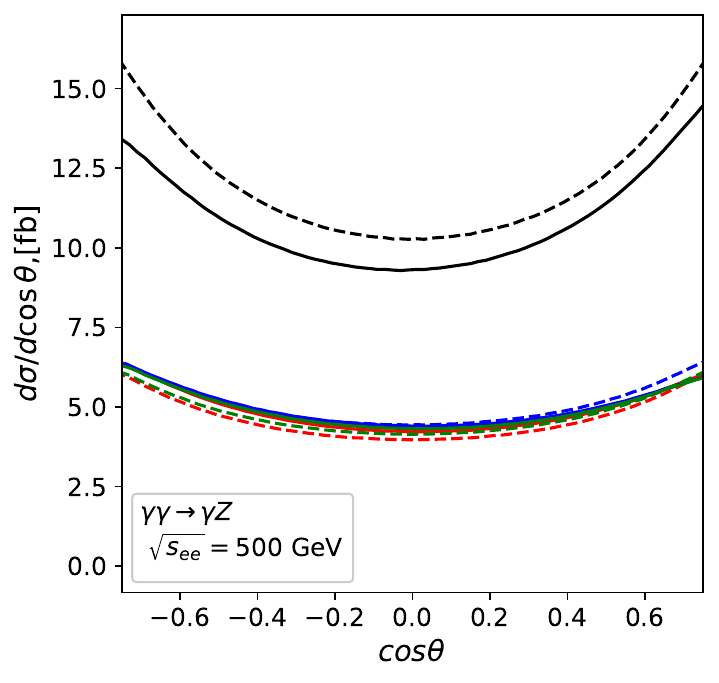}
\includegraphics[width=0.315\textwidth]{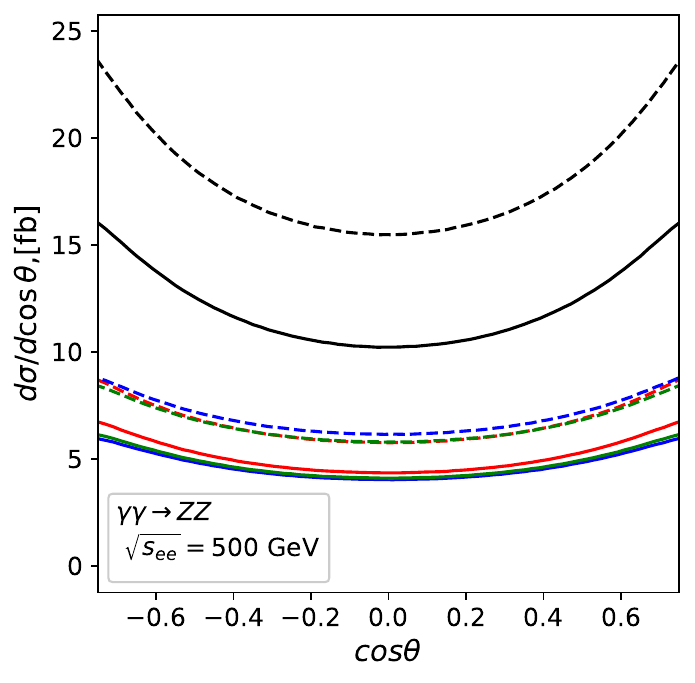} 
  \end{center}
  \caption{Final state $\cos\theta$ kinematic distributions for electron beam energy
  $\sqrt{s_{ee}} = 500$~GeV and various polarizations (see text). The color legend
  for $\gggg$ plot applies to the other two.\label{fig:kin_cth3}}
\end{figure*}
\begin{figure*}
  \begin{center}
\includegraphics[width=0.315\textwidth]{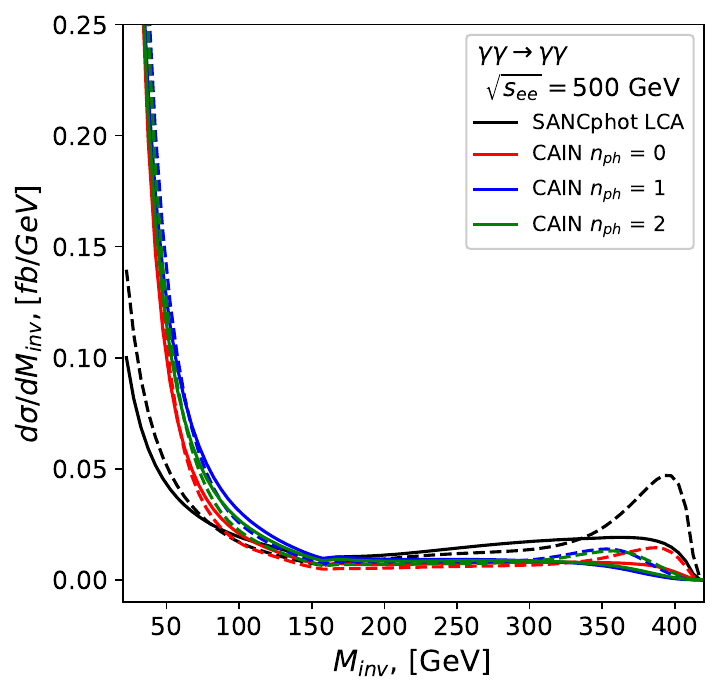}
\includegraphics[width=0.315\textwidth]{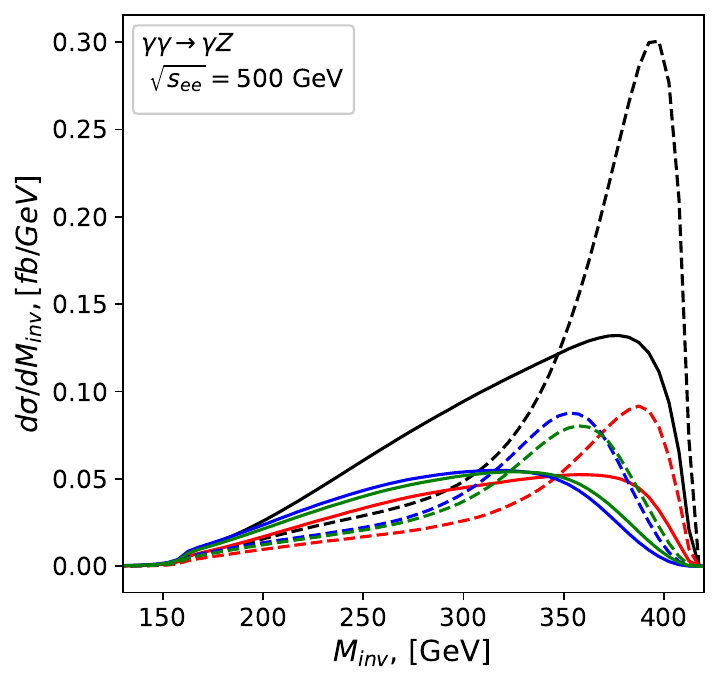}
\includegraphics[width=0.315\textwidth]{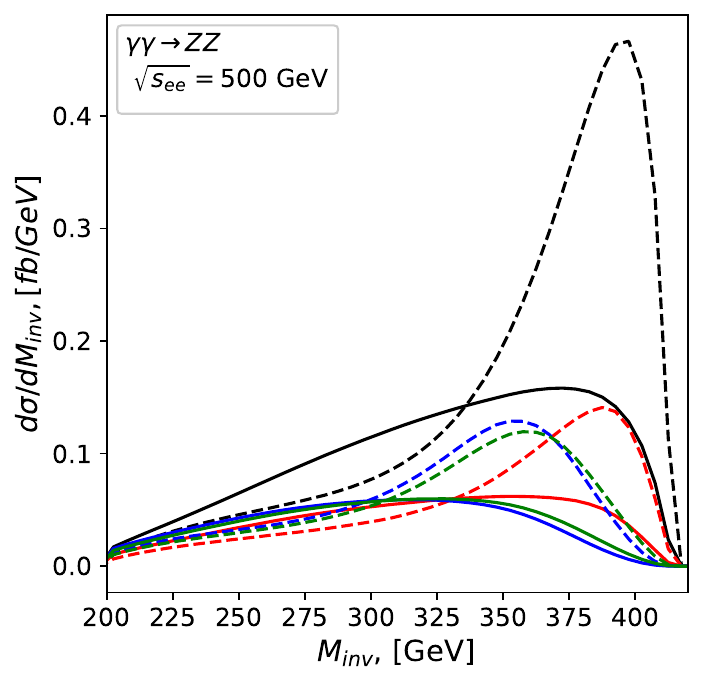} \\
\includegraphics[width=0.315\textwidth]{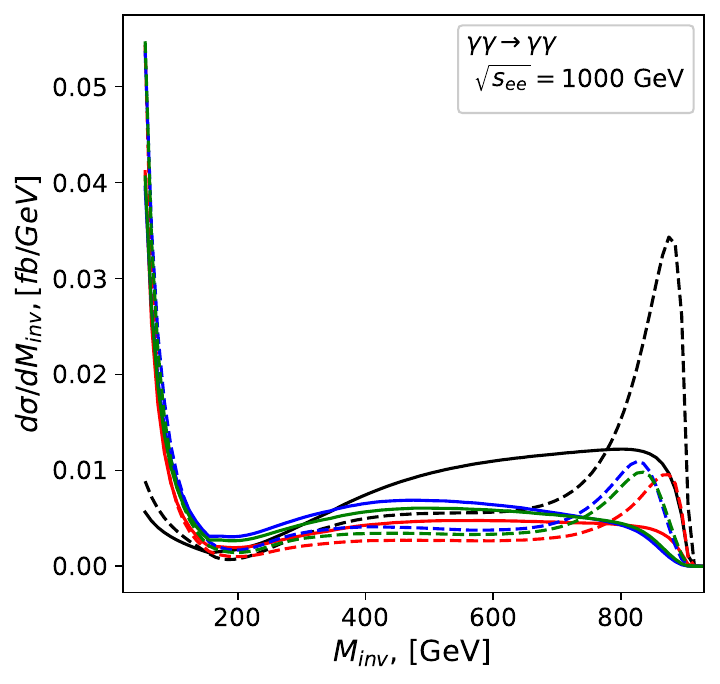}
\includegraphics[width=0.315\textwidth]{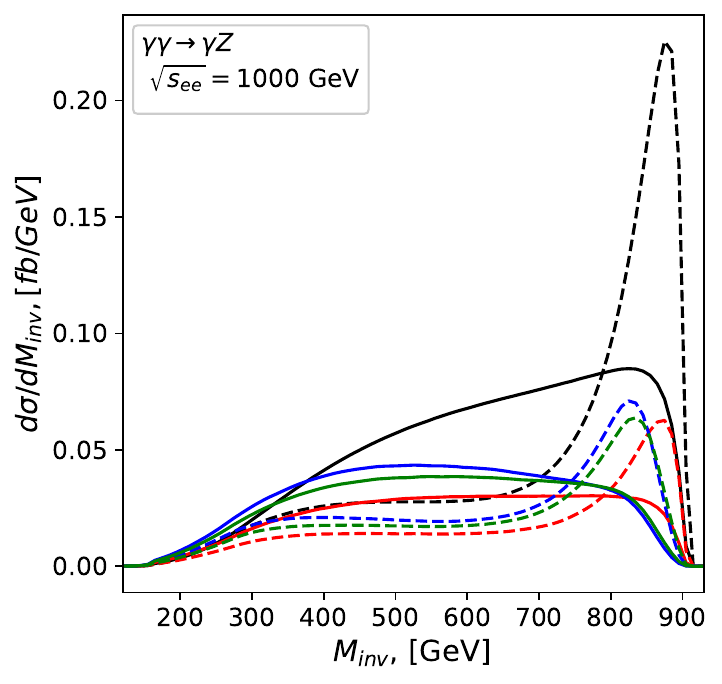}
\includegraphics[width=0.315\textwidth]{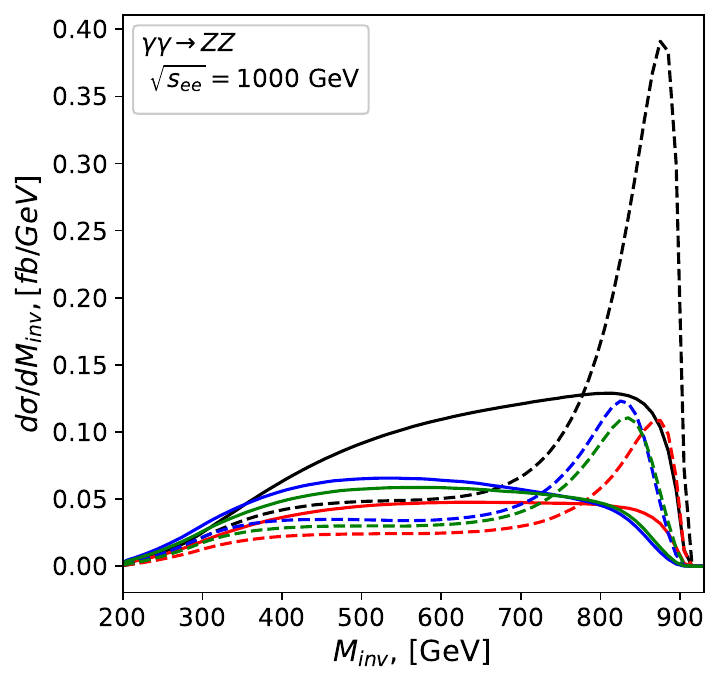}
  \end{center}
  \caption{Final state invariant mass kinematic distributions for different electron beam energies 
  $\sqrt{s_{ee}} = 500$~GeV (upper row) and 1000~GeV (lower row)
  and various polarizations (see text). The color legend
  for 500~GeV $\gggg$ plot applies to the others.\label{fig:kin_m34}}
\end{figure*}
In the invariant mass distributions, we observe a well-known energy-dampening effect of taking
into account realistic beam geometry. The shift to the less energetic area of spectra is
mostly visible for the $\gggg$ process, where it also leads to an increase in the integrated
cross sections due to higher statistics in the central region (Figure~\ref{fig:kin_cth3}).
For the other processes $\gggz$ and $\ggzz$, the lower energy region is kinematically forbidden,
and the total cross section for CAIN-based spectra is reduced compared to the LCA.
\begin{figure*}
  \begin{center}
\includegraphics[width=0.315\textwidth]{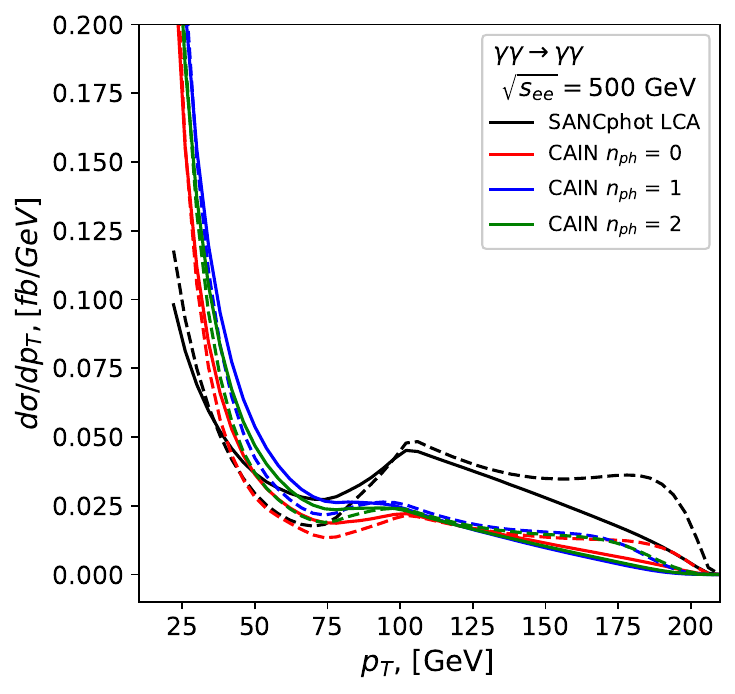}
\includegraphics[width=0.315\textwidth]{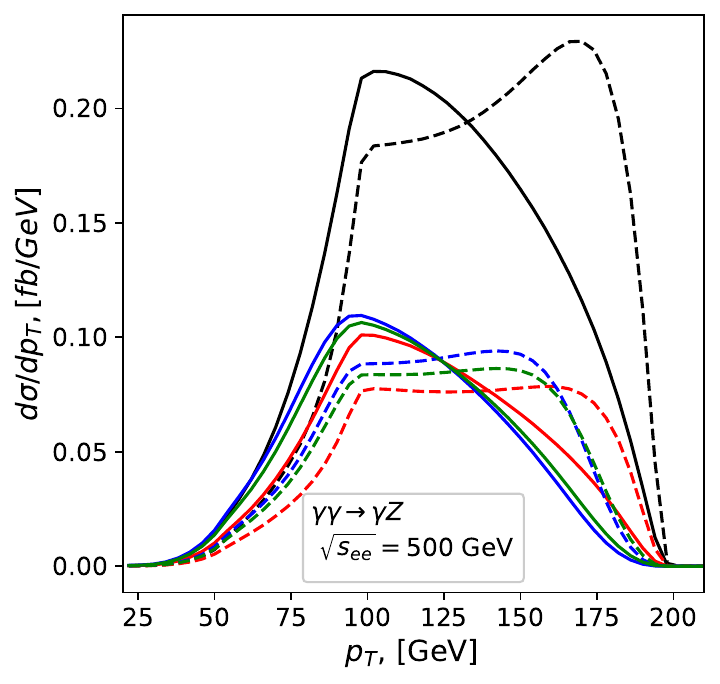}
\includegraphics[width=0.315\textwidth]{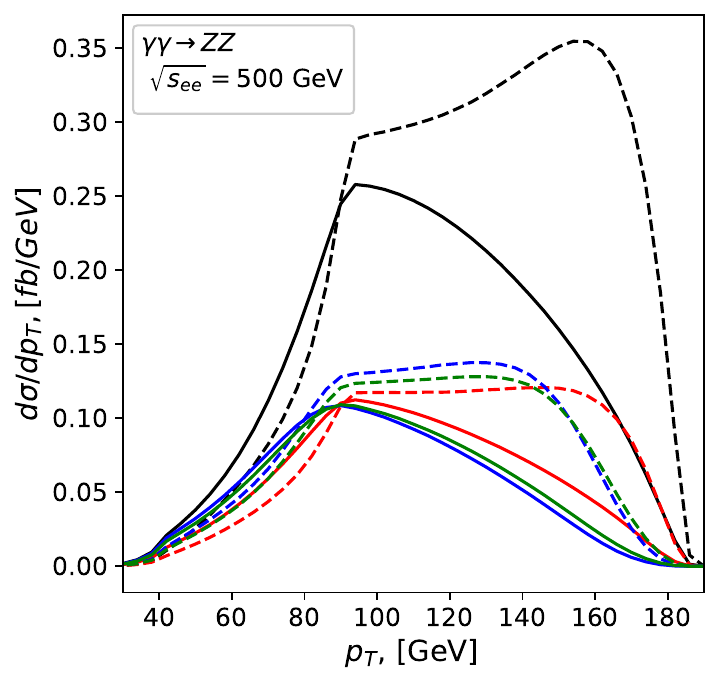} \\
\includegraphics[width=0.315\textwidth]{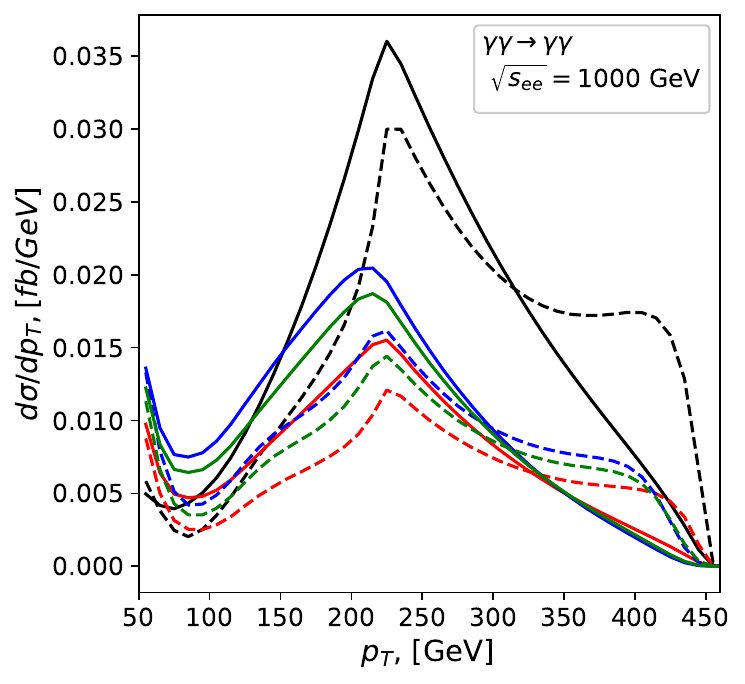}
\includegraphics[width=0.315\textwidth]{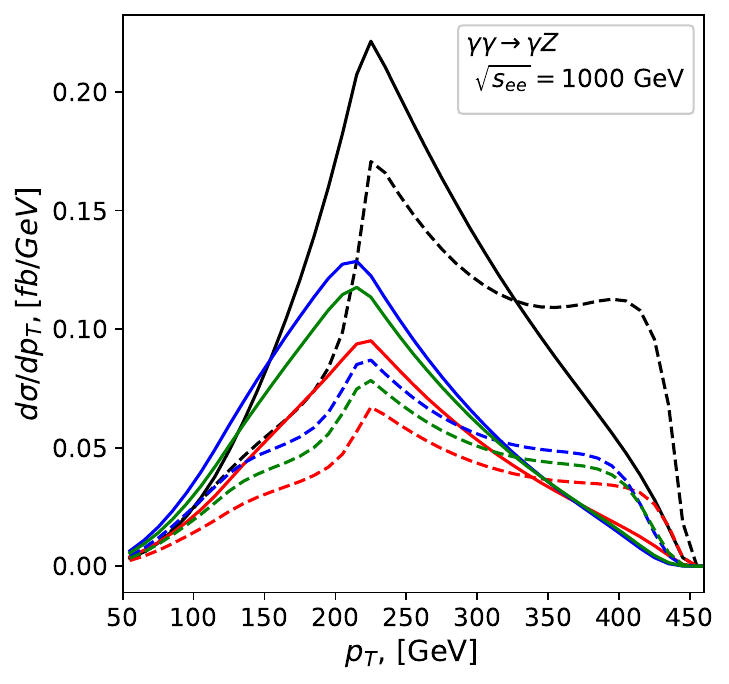}
\includegraphics[width=0.315\textwidth]{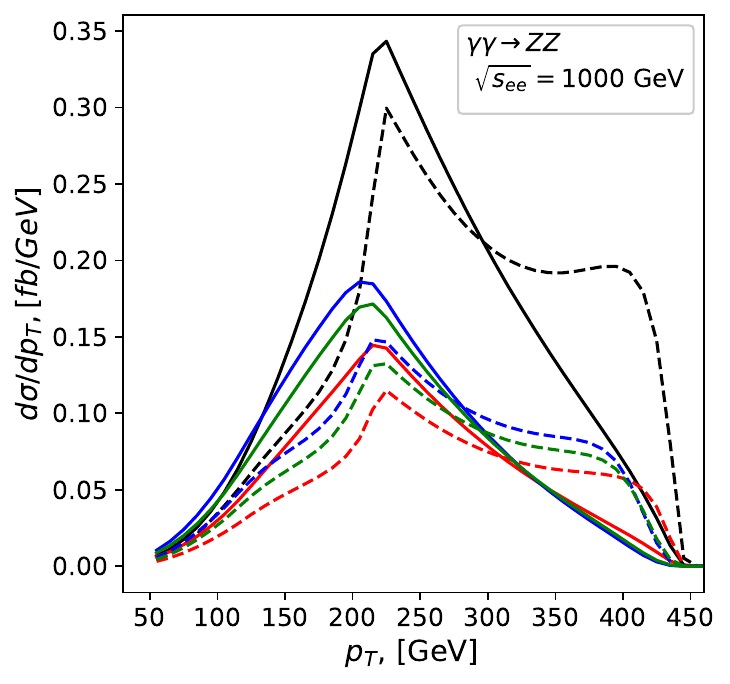} \\
  \end{center}
  \caption{Final state transverse momentum kinematic distributions for different electron beam energies 
  $\sqrt{s_{ee}} = 500$~GeV (upper row) and 1000~GeV (lower row)
  and various polarizations (see text). The color legend
  for 500~GeV $\gggg$ plot applies to the others.\label{fig:kin_pt3}}
\end{figure*}
Another interesting observation on the Figures~\ref{fig:kin_m34} and~\ref{fig:kin_pt3} is that 
the actual effect of the Compton non-linearity degree becomes more prominent at higher energies and
is more noticeable on $p_T$ distributions. Such an effect is likely attributed to the fact that  
this observable is not Lorentz-invariant.
\begin{table*}
\centering
\begin{tabular}{llrrr}
\toprule
 \multicolumn{2}{l}{$n_{ph}$} &  0 (LCA) & 1 & 2    \\
\midrule
$\gamma\gamma \rightarrow \gamma\gamma$, & CAIN  & 13.8457(7)&14.8800(7)&14.3350(7) \\
\cmidrule{2-5}
500 GeV & SANCphot  & 8.0457(4)&-&- \\
\midrule
$\gamma\gamma \rightarrow \gamma\gamma$, & CAIN  & 10.3750(5)&8.9332(4)&9.3450(5) \\
\cmidrule{2-5}
1000 GeV & SANCphot  & 6.8598(3)&-&- \\
\midrule
$\gamma\gamma \rightarrow \gamma Z$, & CAIN  & 8.7229(4)&8.8690(4)&8.7565(4) \\
\cmidrule{2-5}
500 GeV & SANCphot  & 19.3739(1)&-&- \\
\midrule
$\gamma\gamma \rightarrow \gamma Z$, & CAIN  & 17.3690(9)&23.014(1)&21.136(1) \\
\cmidrule{2-5}
1000 GeV & SANCphot  & 38.279(2)&-&- \\
\midrule
$\gamma\gamma \rightarrow ZZ$, & CAIN  & 9.2385(5)&8.3770(4)&8.5766(4) \\
\cmidrule{2-5}
500 GeV & SANCphot  & 21.809(1)&-&- \\
\midrule
$\gamma\gamma \rightarrow ZZ$, & CAIN  & 25.644(1)&32.491(2)&30.055(1) \\
\cmidrule{2-5}
1000 GeV & SANCphot  & 58.945(3)&-&- \\
\bottomrule

\end{tabular}
  \caption{Integrated cross sections $\sigma(\gamma\gamma)$~[fb] with different orders $n_{ph}$ for linear polarization.\label{tab:xs_linear}}
\end{table*}

\begin{table*}
\centering
\begin{tabular}{llrrr}
\toprule
 \multicolumn{2}{l}{$n_{ph}$} &  0 (LCA) & 1 & 2    \\
\midrule
$\gamma\gamma \rightarrow \gamma\gamma$, & CAIN  & 16.2857(8)&17.0891(9)&16.4330(8) \\
\cmidrule{2-5}
500 GeV & SANCphot  & 9.4432(5)&-&- \\
\midrule
$\gamma\gamma \rightarrow \gamma\gamma$, & CAIN  & 13.2011(7)&10.6979(5)&11.6521(6) \\
\cmidrule{2-5}
1000 GeV & SANCphot  & 6.9403(3)&-&- \\
\midrule
$\gamma\gamma \rightarrow \gamma Z$, & CAIN  & 8.2950(4)&9.0302(4)&8.5068(4) \\
\cmidrule{2-5}
500 GeV & SANCphot  & 21.528(1)&-&- \\
\midrule
$\gamma\gamma \rightarrow \gamma Z$, & CAIN  & 13.6011(7)&18.2681(9)&16.1505(8) \\
\cmidrule{2-5}
1000 GeV & SANCphot  & 35.158(2)&-&- \\
\midrule
$\gamma\gamma \rightarrow ZZ$, & CAIN  & 12.0139(6)&12.4726(6)&11.8503(6) \\
\cmidrule{2-5}
500 GeV & SANCphot  & 32.335(2)&-&- \\
\midrule
$\gamma\gamma \rightarrow ZZ$, & CAIN  & 22.620(1)&30.058(1)&26.679(1) \\
\cmidrule{2-5}
1000 GeV & SANCphot  & 59.679(3)&-&- \\
\bottomrule

\end{tabular}
  \caption{Integrated cross sections $\sigma(\gamma\gamma)$~[fb] with different orders $n_{ph}$ for circular polarization.\label{tab:xs_circular}}
\end{table*}

From the total cross section values, we see that for the $\gggg$ process, adding a
realistic beam geometry to the photon spectrum simulation increases the
integrated cross section in the given calculation limits by a factor of 2
depending on whether the polarization is linear or circular. However, this effect
is washed away for higher collision energies and other processes, where, due to the
kinematic threshold, the effect becomes reversed, and for 500~GeV energy, the total
cross section for the $\gggz$ process becomes 3 times less after the inclusion of
non-linear effects and beam geometry.

\begin{figure*}
  \begin{center}
\includegraphics[width=0.3\textwidth]{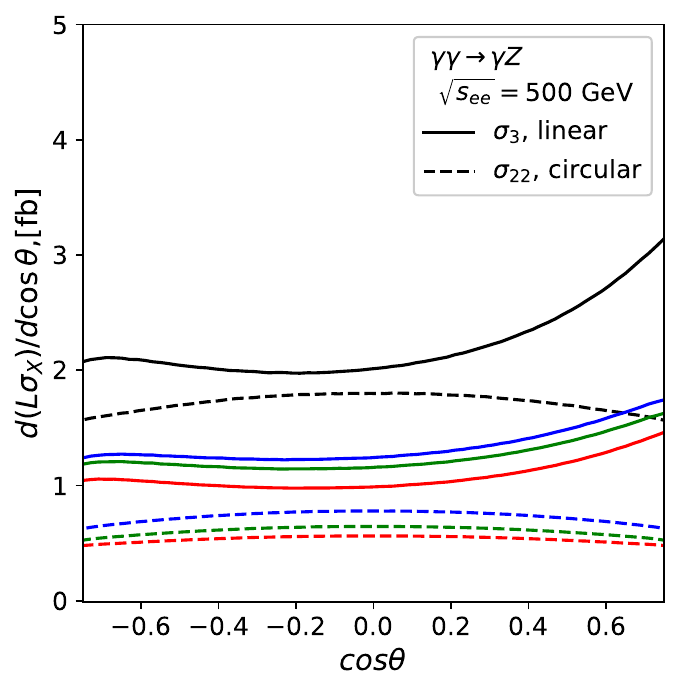}
\includegraphics[width=0.325\textwidth]{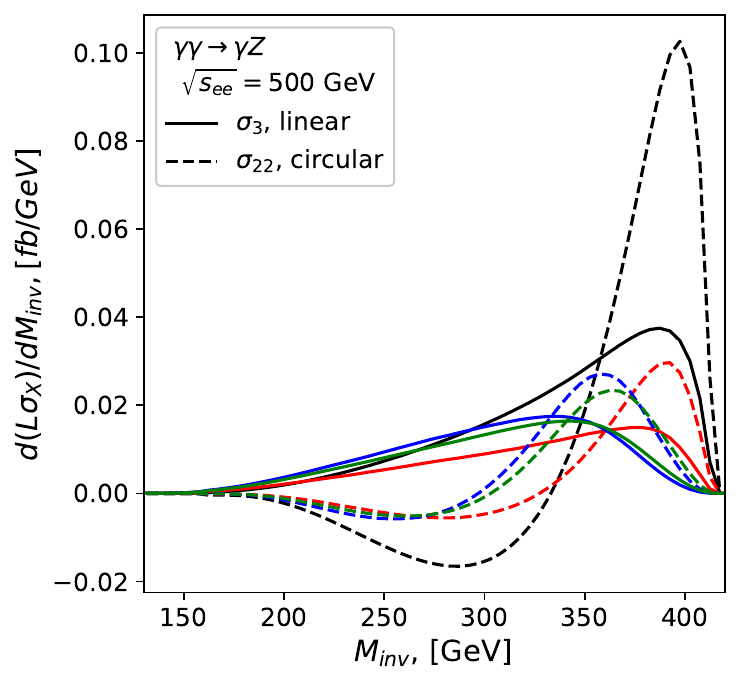}
\includegraphics[width=0.325\textwidth]{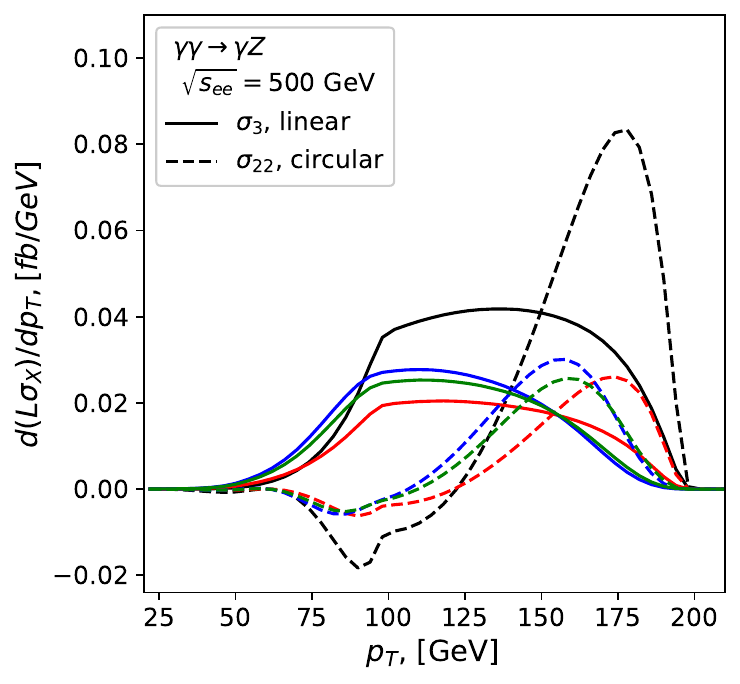} \\
  \end{center}
  \caption{Contributions of $\sigma_{22}$ and $\sigma_3$ to the kinematic distributions 
  for $\gggz$ process at $\sqrt{s_{ee}} = 500$~GeV. The line colors' meaning
  is the same as in Figures~\ref{fig:kin_m34} and~\ref{fig:kin_pt3}.\label{fig:kin_dist_sc}}
\end{figure*}

Related to the BSM studies, it is also interesting to look at the separate cross
section components, i.e. $\sigma_{22}$ and $\sigma_3$ which are sensitive to
particular contributions of SUSY particles as stated in~\cite{Gounaris:1999gh}.
Figure~\ref{fig:kin_dist_sc} shows that adding a detailed beam geometry
description leads to a dampening of these components, similar to the total cross
section, without particular changes in the distribution shapes. The next
orders of approximation exhibit shifting of the spectra to the lower energies.

Being able to include a more realistic description of the photon spectra is
definitely beneficial for precise studies of $\gamma\gamma$-induced processes.
It would also be interesting to look at the higher orders of Compton
non-linearity for linear polarizations, as this would allow us to test the different
polarization variants depending on the $\phi$ direction. Another fruitful topic
of future research would be adding the decay of the final Z particles and thus
gaining access to a broad range of studies related to polar angle asymmetries.
One more improvement to this study would be the inclusion of the depolarisation
effects into account, which should further improve our precision.

As the input photon energy distribution is agnostic of the photon generation
process, this new approach allows the simulation of the
\texttt{SANCphot} processes with initial photons produced not only via Compton
back-scattering, but also via the beamstrahlung~\cite{Telnov:1989sd, Yokoya:1991qz} given the 
spectra, generated, for example, with GuineaPig program~\cite{Schulte:1998au}.

\section{Summary}
\label{sec:summary}
Thanks to the inclusion of a piecewise description of general photon spectra
along with the polarization information, the precision of photon-photon
collision simulation in \texttt{SANCphot} is substantially improved. With
realistic photon beam simulation by the CAIN program, we see that the total
cross section of the $\gggg$ process is increased by 50 - 100\% depending on the
beam energy, and for the processes with final Z boson(s) is reduced by a factor
of 2. In addition, the kinematic distributions show a noticeable shift to lower
energies while keeping the polarization distinctive features intact.

\section{Acknowledgements}
We thank V.~Telnov and I.~Ginzburg for their advice and fruitful discussion.

\section*{Funding}

This research has been funded by the Committee of Science of the Ministry
of Science and Higher Education of the Republic of Kazakhstan (Grant
No. AP19680084).


\bibliography{main}

\end{document}